# Shubnikov–de Haas oscillations and planar Hall effect in HfTe$_2$

Qixuan Li,[1] Gangjian Jin,[2] Nannan Tang,[1] Bin Wang,[1] Bing Shen,[1] Donghui Guo,[1] Dingyong Zhong,[3] Huakun Zuo,[2,*] and Huichao Wang[1,*]

[1]*Guangdong Provincial Key Laboratory of Magnetoelectric Physics and Devices, School of Physics, Sun Yat-sen University, Guangzhou 510275, China*

[2]*Wuhan National High Magnetic Field Center and School of Physics, Huazhong University of Science and Technology, Wuhan 430074, China*

[3]*State Key Laboratory of Optoelectronic Materials and Technologies, School of Physics, Sun Yat-sen University, Guangzhou 510275, China*

**Abstract**

Layered transition-metal dichalcogenide (TMD) HfTe$_2$ is a topological semimetal candidate with increasing attentions recently. The map of the Fermi surface is of interest and importance to understand its properties. Here we present a study of Shubnikov–de Haas (SdH) oscillations and planar Hall effect (PHE) in HfTe$_2$. The single crystals grown by flux method show the largest unsaturated magnetoresistance (MR) effect of $1.1\times10^4$ % at 14 T and 2 K. The angle-resolved SdH oscillations reveal that the Fermi surface consists of three pockets with different anisotropy. In addition, we observe PHE and anisotropic MR (AMR) effect in the material for a wide temperature range. The effective mass, carrier density and quantum transport mobility are quantified in the system, and the Berry phase is discussed. Our work provides crucial insights into the electronic structure and the Fermi surface of the semimetal.

## I. INTRODUCTION

Transition-metal dichalcogenides (TMDs) are a kind of materials attracting broad interest because of their rich physical properties and promising applications. The TMDs with nontrivial electronic structure have been a recent topic and shown various novel transport behavior including remarkable unsaturated MR, chiral anomaly, temperature-induced Lifshitz transition and PHE [1-11]. These fantastic phenomena have triggered intensive studies on a variety of TMDs [12-18].

Hafnium ditelluride (HfTe$_2$) crystallized in the 1T structure has been recently predicted to be a topological semimetal candidate [19-22]. The first-principles band-structure calculations reveal that bulk HfTe$_2$ shows two hole pockets crossing the Fermi level at the Γ point and one electron pocket around the Brillouin-zone corner [23]. The low-temperature angle-resolved photoemission spectroscopy (ARPES) experiments confirm the existence of multiple bands and explore the occurrence of a bulk Dirac point [22-24]. The magnetization measurements reveal one hole pocket and one electron pocket from de Haas-van Alphen effect [25]. However, the information of another hole Fermi pocket is lacking and it remains a question whether it is an open- orbit Fermi surface as theoretically predicted [23,25]. The angular-dependent SdH effect in MR has been an important tool to map the Fermi surface but there has been few experimental work to show the electronic transport behavior of HfTe$_2$, which may be owing to the challenge of growing high quality single crystals and the extreme sensitivity of HfTe$_2$ in atmosphere. Thus, the electronic transport measurements of high-mobility HfTe$_2$ single crystals are highly desirable for understanding the electronic structure and transport properties.

Here we present the SdH quantum oscillations and PHE in high quality HfTe$_2$ single crystals. The crystals were grown by self-Te-flux method to avoid introducing other impurities. The best crystal shows a residual resistance ratio (*RRR*) of 234 and an unsaturated MR effect of $1.1\times10^4$ % at 14 T and 2 K. We observe obvious SdH oscillations in the MR curves at low temperatures, which reveal three Fermi pockets consistent with the theoretical prediction. The three pockets show different electronic anisotropy identified by the angle-resolved SdH effect. PHE was observed in the material and may result from the nonzero Berry phase and the in-plane AMR. This work provides insights into the transport properties supporting the current understanding of HfTe$_2$ electronic structure.

## II. EXPERIMENTAL METHODS

TABLE 1. Growth parameters and physical properties of HfTe$_2$ single crystals grown by self-Te-flux method. The samples are represented by A-D denoting different Hf:Te ratios in initial ingredients and numbers to distinguish each other. The *RRR* is calculated by $\rho(300K)/\rho(2K)$ and the MR, *i.e.*, $(R(H)-R(0))/R(0)$ is measured at 2 K and 14 T in PPMS. More than thirty samples from different batches are measured, and SdH oscillations with three Fermi pockets are observed in all series of samples. Typical samples are selected to show the properties.

| Sample | Hf:Te | *RRR* | MR($\times 10^3$ %, 2K, 14T) |
|---|---|---|---|
| A1 | 4.5 : 95.5 | 234.2 | 10.77 |
| A2 | 4.5 : 95.5 | 203.6 | 7.17 |
| A4 | 4.5 : 95.5 | 153.39 | 5.54 |
| A5 | 4.5 : 95.5 | 167.77 | 7.65 |
| B1 | 5 : 95 | 211.6 | 7.54 |
| B2 | 5 : 95 | 200.5 | 7.29 |
| C1 | 4 : 96 | 215.8 | 9.39 |
| C2 | 4 : 96 | 197.5 | 7.89 |
| D1 | 3.5 : 96.5 | 194.4 | 6.85 |
| D2 | 3.5 : 96.5 | 179.8 | 7.73 |

HfTe$_2$ is trigonal omega structured (space group $P\bar{3}m1$) with HfTe$_2$ layers stacked along the (001) direction. Hf$^{4+}$ is bonded to six equivalent Te$^{2-}$ atoms to form edge-sharing HfTe$_6$ octahedra. The HfTe$_2$ can be synthesized via chemical vapor transport using iodine as a transport agent and this method is widely used in previous studies [23-26]. The self-Te-flux growth method appears to be superior to yield high quality crystals owing to the absence of a transport agent which may cause defects and impurities [27]. We have grown HfTe$_2$ single crystals by a Te-rich melt, i.e., Te as the flux in the process. Total elements mass of 5 g (Hf, 99.99%, Zr < 0.03%; Te, 99.9999%) were loaded in a quartz tube with four different ratios (Hf:Te=3.5:96.5, 4:96, 4.5:95.5, 5:95). Tubes were then sealed (pressure < $5\times10^{-4}$ mbar), heated to 1000 °C within 16 h, held at 1000 °C (24 h), and cooled to 700 °C (rate of 2 °C/h) to remove the flux by a centrifuge. The as-synthesized crystals were characterized by energy dispersive X-ray spectroscopy (EVO MA10) to show a Hf:Te ratio close to 1:2 [28]. The HfTe$_2$ single crystal is intrinsically gold in color [28] while it is extremely sensitive in the air and easily oxidizes to form a dark grey film on the surface. Therefore, the crystals were kept in the glove box before measurements for protection. The transport measurements were performed in a 14T-DynaCool PPMS (Physical Property Measurement System, Quantum Design). Silver glue was used to paste the gold wires to the crystals after obtaining the golden fresh surfaces by scraping off the top oxidized films. Standard four-probe or six-probe method was used to measure the MR and the Hall behavior.

## III. RESULTS AND DISCUSSION

The typical zero-field longitudinal resistivity of a HfTe$_2$ crystal as a function of temperature is shown in Fig. 1(a). It shows a transition from linear behavior $\rho \sim T$ at high temperature to $\rho \sim T^3$ at low temperature [Fig. 1(a) inset]. It is known that the linear temperature dependence originates from the electron-phonon coupling. The $T^3$ dependence of resistivity is different from the expected $T^2$ behavior of a Fermi liquid state with dominant electron-electron scattering at low temperature. It is related to the *s-d* interband scattering and generally observed in transition-metal compounds [29]. The residual resistivity ratio $RRR=\rho(300K)/\rho(2K)$ is an important indicator of the crystal quality and purity. The $RRR$ of high quality $HfTe_2$ crystals synthesized via self-Te-flux method can be as high as 404 [30]. The largest $RRR$ in our obtained samples is 234 (sample A1). The crystal also shows the largest MR=$[\rho(H)-\rho(0)]/\rho(0)$ of about $1.1\times10^4$ % at 14 T and 2 K [Fig. 1(b)]. We measured the RT and MR behavior of a variety of samples from different batches [TABLE 1]. In a fixed magnetic field, the MR is largely influenced by the carrier mobility $\mu$ which is proportional to the conductivity $\sigma_0$ at low temperature. Meanwhile, the $RRR$ is proportional to the conductivity $\sigma_0=1/\rho_0$. Thus, for different batches of $HfTe_2$ samples, the MR is positively correlated with the $RRR$ as observed [Fig. 1(c)].

The Hall measurements are performed to analyze the carrier information in the material. Based on the measured $\rho_{xx}$ [Fig. 1(d)] and $\rho_{yx}$ [Fig. 1(e)], the Hall conductivity $\sigma_{xy}$ can be calculated by $\sigma_{xy}=\rho_{yx}/(\rho_{xx}^2+\rho_{yx}^2)$ [Fig. 1(f)]. In order to compare with the results from quantum oscillations discussed later, the Hall conductivity is fitted using a three-band model. The obtained carrier densities at 2 K are $n_{h1}=6.6\times10^{19}$ cm$^{-3}$, $n_e=3.9\times10^{19}$ cm$^{-3}$ and $n_{h2}=8.7\times10^{19}$ cm$^{-3}$, respectively. The Hall analysis reveals two hole pockets and an electron pocket, consistent with the first-principle calculations and the ARPES experiments [23-25]. In addition, the Hall mobilities at 2 K are estimated to be $\mu_{h1}=6.3\times10^3$ cm$^2$V$^{-1}$s$^{-1}$, $\mu_e=5.1\times10^3$ cm$^2$V$^{-1}$s$^{-1}$ and $\mu_{h2}=7.7\times10^3$ cm$^2$V$^{-1}$s$^{-1}$, respectively.

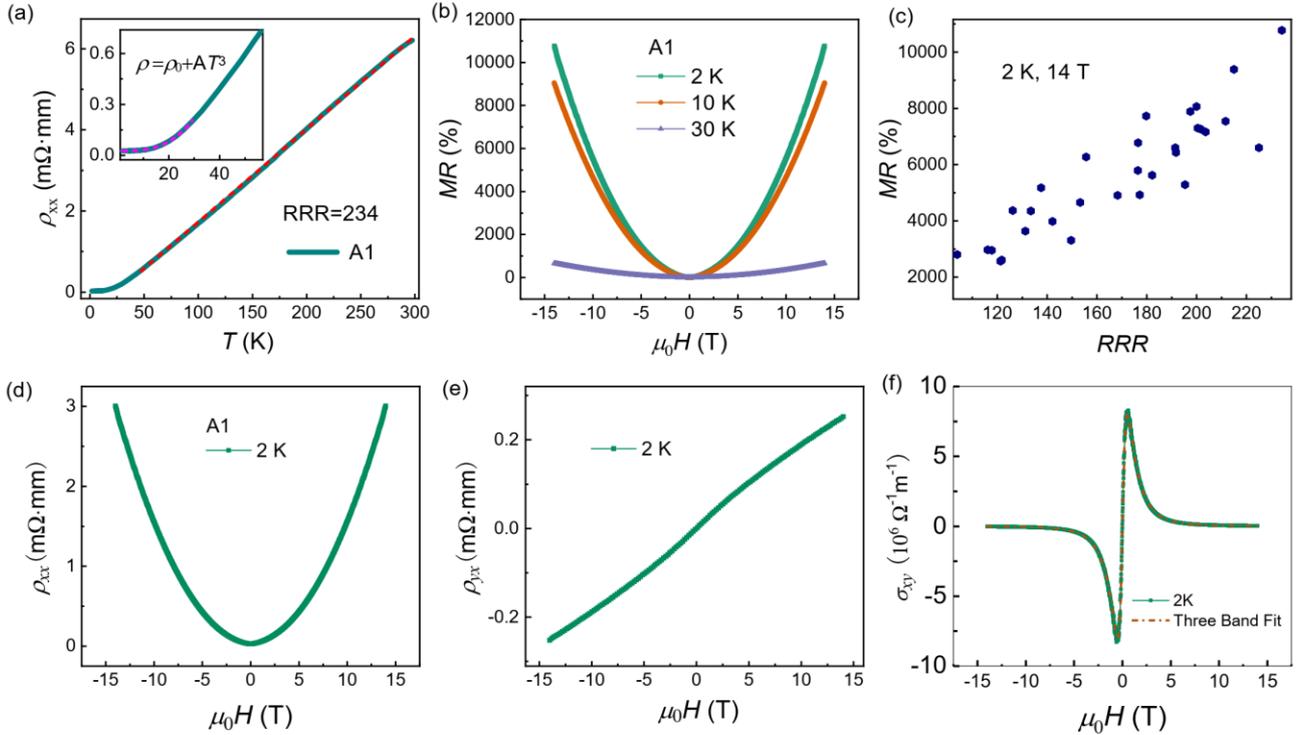

FIG. 1. (a) Temperature dependence of zero-field resistivity (sample A1). Inset: Low temperature resistivity shows $\rho\sim T^3$ behavior indicating dominant s-d interband scattering. (b) MR curves of $HfTe_2$ at low temperature. (c) The MR is positively correlated with the $RRR$. (d) Magnetoresistivity $\rho_{xx}$ at 2 K. (e) Hall resistivity $\rho_{yx}$ at 2 K. (f) Carrier concentration and mobility estimated from the three-band model fitting of Hall conductivity $\sigma_{xy}$.

It is observed that there are oscillations on the low temperature MR curves in the high magnetic field range of 8-14 T [Fig. 2(a)]. The oscillatory component $\Delta R$ [Fig. 2(b)] is extracted by the commonly used method that subtracting a smooth quadratic polynomial background from data in Fig. 2(a) considering the parabolic shape of the MR curve. We have also used different orders of polynomial as the background and performed the second derivative of the MR data (Fig. S1 of Supplemental Material) [31], confirming the quantum oscillations are intrinsic. The direct fast Fourier transformation (FFT) [Fig. 2(c)] of the oscillations with respect to $1/\mu_0H$ reveals prominent frequency peaks locating at around 131 T ($\alpha$), 262 T ($2\alpha$), 355 T ($\beta$), 430 T ($\gamma$), 710 T ($2\beta$) and 970 T ($2\alpha+2\beta$), respectively. The peaks of 262 T and 710 T are the second harmonic, and the 972 T peak serves as a signature of magnetic breakdown orbits as observed in the semimetal $WTe_2$ [32]. Thus, the SdH oscillations identify three Fermi pockets ($\alpha,\beta,\gamma$), among which the two smaller oscillation frequencies ($\alpha$ and $\beta$) are in agreement with the results detected by the magnetization measurements [25].

In this work, the samples are represented by A-D series denoting different Hf:Te ratios in initial ingredients and number to distinguish each other. According to our measurements on more than thirty samples (Fig. S2 of Supplemental Material), the three FFT frequencies are present in all the series of samples though the $\gamma$ frequency peak is very weak in some samples measured within 14 T and the noise increases the difficulty in identifying it [31]. It is noted that the FFT peak amplitudes of the three pockets are different, which can be attributed to the differences of Dingle temperature $T_D$ and effective mass $m^*$ for the three Fermi pockets as analyzed below. In different samples, the difference of disorder level can lead to the variance of the FFT peak amplitudes because the Dingle temperature is closely related to the electron scattering, and the disorder induced slight variance of the Fermi level may make the effective mass $m^*$ different.

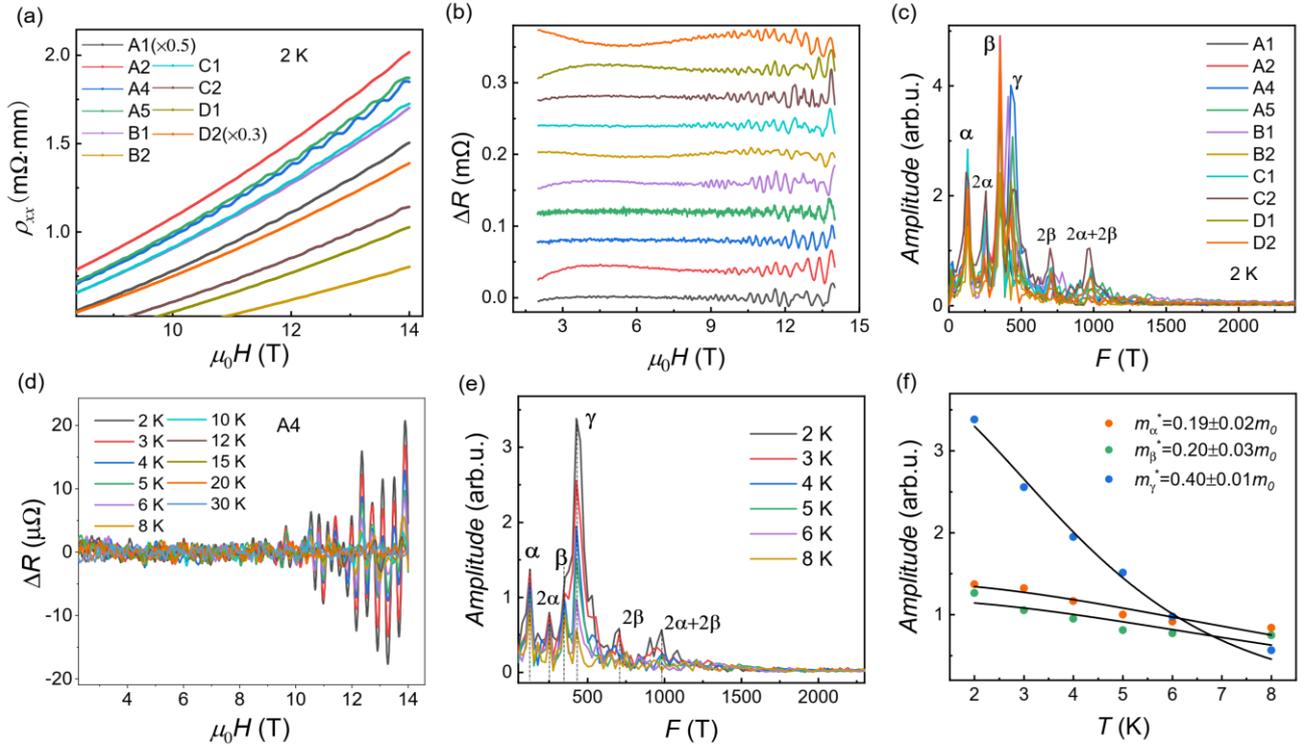

FIG. 2. (a) MR curves of different samples with oscillations in the high magnetic field regime at 2 K. (b) The extracted MR oscillations after subtracting a smooth quadratic polynomial background. The plots are shifted vertically for clarity. (c) The FFT results of the oscillations reveal three Fermi pockets in the material. (d) The MR oscillations at selected temperatures (sample A4). (e) The temperature dependence of the FFT spectrum in sample A4. (f) The effective mass fits (black lines). The points are the FFT peak amplitudes at different temperatures.

According to the Lifshitz-Onsager quantization rule, the quantum oscillations frequency ($F$) is proportional to the extremal cross-sectional areas $S_F$ of the Fermi surface $F=(\hbar/2\pi e)S_F$. By assuming a circular Fermi pocket, the Fermi vector can be estimated based on $S_F= \pi k_F^2$. Then the three-dimensional (3D) carrier concentrations can be extracted with $n=k_F^3/3\pi^2$ using the spherical approximation [33]. The estimated carrier concentrations for $\alpha$, $\beta$ and $\gamma$ pockets of sample A4 are $8.5\times10^{18}$ cm$^{-3}$, $3.8\times10^{19}$ cm$^{-3}$, $5.0\times10^{19}$ cm$^{-3}$, respectively. The results are close to those obtained from Hall measurement. The differences may be attributed to that we did not consider the anisotropy of the Fermi surface in the estimation of carrier densities based on the SdH oscillation frequency.

The SdH oscillations of a 3D system are described by the standard Lifshitz-Kosevich (LK) theory [34-37]:

$$\Delta R_{xx} \propto \exp(\frac{-2\pi^2 k_B T_D}{\hbar \omega_c}) \frac{\frac{2\pi^2 k_B T}{\hbar \omega_c}}{\sinh\left(\frac{2\pi^2 k_B T}{\hbar \omega_c}\right)} \cos 2\pi(\frac{F}{B} + \phi)$$

where $T_D$ is the Dingle temperature, $\omega_c$ is the cyclotron frequency related to the effective mass $m^*$ by $\omega_c = e\mu_0 H/m^*$, $k_B$ is Boltzmann's constant, $\hbar$ is Planck's constant divided by $2\pi$, and $\phi$ is the oscillating phase. The temperature dependent SdH oscillations [Fig. 2(d)] are obtained from the measured MR behavior of sample A4 at different temperatures (Fig. S3 of Supplemental Material). It is observed that the oscillations are absent at higher temperatures [31]. The FFT results at low temperatures are shown in Fig. 2(e). The SdH oscillations of a system with multiple Fermi pockets are described by linear superposition of the LK formula for each frequency and the amplitude of each FFT frequency peak depends on the amplitude of MR oscillations from each Fermi pocket. The FFT amplitude follows FFT$_{ampl}$ ~ $\lambda(T)/\sinh\lambda(T)$ with $\lambda(T)=2\pi^2 k_B T m^*/\hbar e\mu_0 \overline{H}$ where $\overline{H}$ is the average magnetic field for the applied magnetic field range [38]. In this case, the temperature dependent amplitude of each FFT peak can be used to extract the corresponding effective (cyclotron) mass of the carriers. This method is widely used in various systems including the semimetals with multiple Fermi pockets [39-42]. As show in Fig. 2(f), the estimated effective cyclotron mass in HfTe$_2$ is $m_\alpha^*=(0.19\pm0.02)m_0$, $m_\beta^*=(0.20\pm0.03)m_0$ and $m_\gamma^*=(0.40\pm0.01)m_0$, respectively. Here $m_0$ represents the bare electron mass. Similar results are obtained for other samples (Figs. S4 and S5) [31].

The SdH oscillations are further verified by the high magnetic field measurements up to 56 T. The newly grown samples with longer size and thinner thickness are chosen for the pulsed magnetic field measurements which show relatively larger noise owing to the mechanical vibrating and need a large signal to obtain reliable data. In Fig. 3(a), the MR data of sample A5 show remarkable oscillations which are extracted to be shown in Fig. 3(b). The FFT results [Fig. 3(c)] reveal three frequency peaks at around 131 T ($\alpha$), 351 T ($\beta$) and 438 T ($\gamma$), respectively. The second harmonic $2\alpha$ and $2\beta$ peaks are also observed with the appearance of $2(\alpha+\beta)$ orbit suggesting the occurrence of magnetic breakdown. Thus, the high field data also identify three Fermi pockets. The fits of the temperature dependent FFT peak amplitudes produce the effective mass $m_\alpha^*=(0.18\pm0.02)m_0$, $m_\beta^*=(0.19\pm0.01)m_0$ and $m_\gamma^*=(0.40\pm0.01)m_0$, respectively [Fig. 3(d)]. The results are confirmed by the high field measurements of another sample (Fig. S6) and coincide with those obtained from the static field data.

The angular dependence of the MR behavior (sample A5) is also measured [Fig. 4(a)]. Here the angle denotes the relative orientation of the magnetic field and the $c$ axis. Hence $\theta=0°$ denotes the configuration that the field is perpendicular to the crystal plane and $\theta=90°$ indicates an in-plane magnetic field perpendicular to the current. The MR at selected angles at 2K can be scaled by the field $\varepsilon_\theta H$ [Fig. 4(b)] with $\varepsilon_\theta=(\cos^2\theta+\eta^2\sin^2\theta)^{1/2}$ where $\eta$ describes the Fermi surface anisotropy [43]. The average value of $\eta$ obtained from several samples is about 6.8 [Fig. 4(c)]. The small $\eta$ reveals the overall 3D electronic structure with anisotropy. The first-principles band-structure calculations predicted that HfTe$_2$ is a semimetal with two hole pockets around the Brillouin-zone center and one electron pocket located around the Brillouin-zone corner. In particular, one of the hole Fermi surface is predicted to be open along $k_z$ forming a warped 2D cylinder [23,25]. This is a question needs to be studied and is important for the understanding of its transport properties.

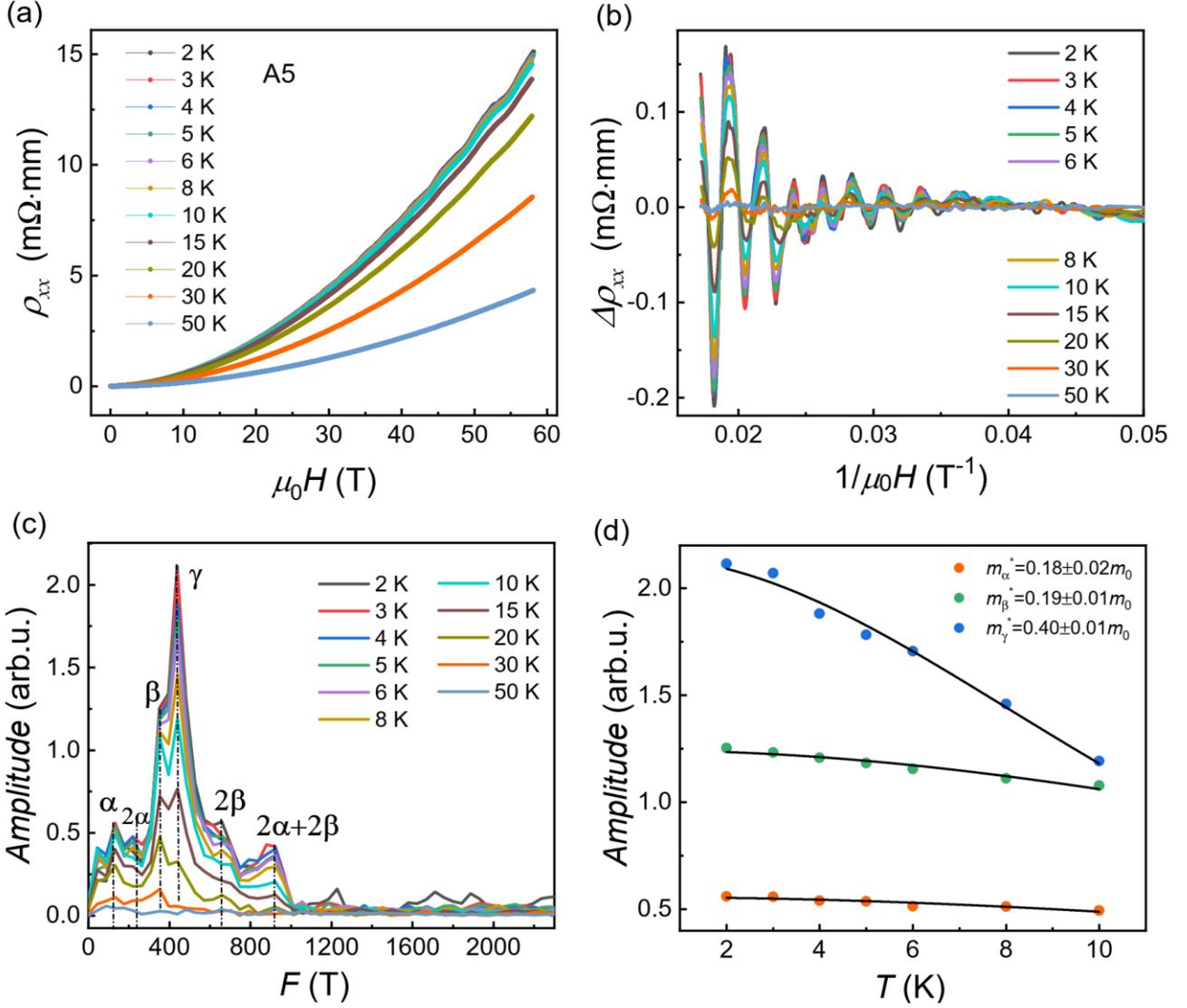

FIG. 3. (a) MR with oscillations at high pulsed magnetic field. (b) The MR oscillations at different temperatures after subtracting a polynomial background. (c) The FFT results. (d) The effective mass fits.

The angular dependence of the SdH oscillations is shown in Fig. 4(d). The angle-dependent FFT results [Fig. 4(e)] show that the three Fermi pockets display different anisotropic properties. The $\beta$ pockets can be recognizable at different angles. The $F_\alpha$ peak is weak at some angles while the peak is robust to appear locating at the frequency range consistent with $\alpha$ pocket. In addition, both the FFT results of the second derivative and polynomial-subtracted oscillation data suggest the emergence of $\alpha$ pocket at different angles (see Figs. S7-S8). Moreover, it is noticed that there are two FFT peaks around 270 T and 610 T when the field is in plane. The frequency peaks cannot be ascribed to $\gamma$ because it is not consistent with the calculated anisotropic feature of the Fermi surface. According to Fig.4(e), the $\gamma$ pocket is visible only when the titled angle is smaller than 60°. Considering the larger frequency of $\gamma$ and thus a larger cross-sectional area of the Fermi surface, the $\gamma$ pocket is suggested to be the predicted hole pocket open in $k_z$ [23,25]. The absence of the $F_\gamma$ peak under an in-plane magnetic field provides support for the judgment since open orbits do not give rise to quantum oscillations.

It is found that the pulsed high magnetic field oscillations in sample B3 are dominated by the $\gamma$ pocket [Figs. S9(a)-(c)], which is demonstrated by the much sharper and larger amplitude of the frequency peak in the FFT spectrum [31]. Especially, the FFT spectrum of oscillations in the field range of 35-56T reveals only one peak of 421 T. The LK formula with the single frequency well reproduces the MR oscillations at large fields [Fig. S9(d)] and the obtained oscillating phase factor $\phi=0.1$. The $\phi$ value close to 0 suggests a nonzero Berry phase close to $\pi$ [31]. It is hard to extract the information of $\alpha$ and $\beta$ pockets from the pulsed high field data owing to the relatively smaller weight in the oscillations and the larger noise while they can be obtained from the fit of the oscillations in the static field data. The fit of the typical oscillations in static magnetic field by LK formula with three different frequencies gives that the oscillating phase factor of $\alpha$ and $\beta$ pockets are 0.76 and 0.40, which is close to the expected value 0.5 of a trivial band with zero Berry phase. The obtained oscillating phase factor of $\gamma$ pocket is 0.04, consistent with the pulsed high magnetic field result suggesting a nonzero $\pi$ Berry phase. The nonzero Berry's phase may arise from a topological Dirac cone. However, the first-principle calculations and the ARPES experiments reveal that the Fermi level of HfTe$_2$ is far away from the predicted Dirac point [23-25], which indicates that the Dirac cone may not play an important role. The band-contact lines usually exist in metals and can also lead to nonzero Berry's Phase [44]. We do not have a clear answer on the origin yet and the physics calls for further investigations. The fitted averaged Dingle temperature $T_D$ of $\alpha$, $\beta$ and $\gamma$ pockets are 16.9 K and 12.3 K and 9.6 K, respectively (Fig. S10) [31]. According to $\mu=e\hbar/2\pi k_B T_D m^*$, the quantum mobility is estimated to be $\mu_\alpha=703$ cm$^2$V$^{-1}$s$^{-1}$, $\mu_\beta=915$ cm$^2$V$^{-1}$s$^{-1}$, $\mu_\gamma=557$ cm$^2$V$^{-1}$s$^{-1}$, respectively. The Hall mobility is much larger than the quantum mobility because the former is only influenced by the large-angle scattering while the latter is sensitive to both small-angle and large-angle scattering [45].

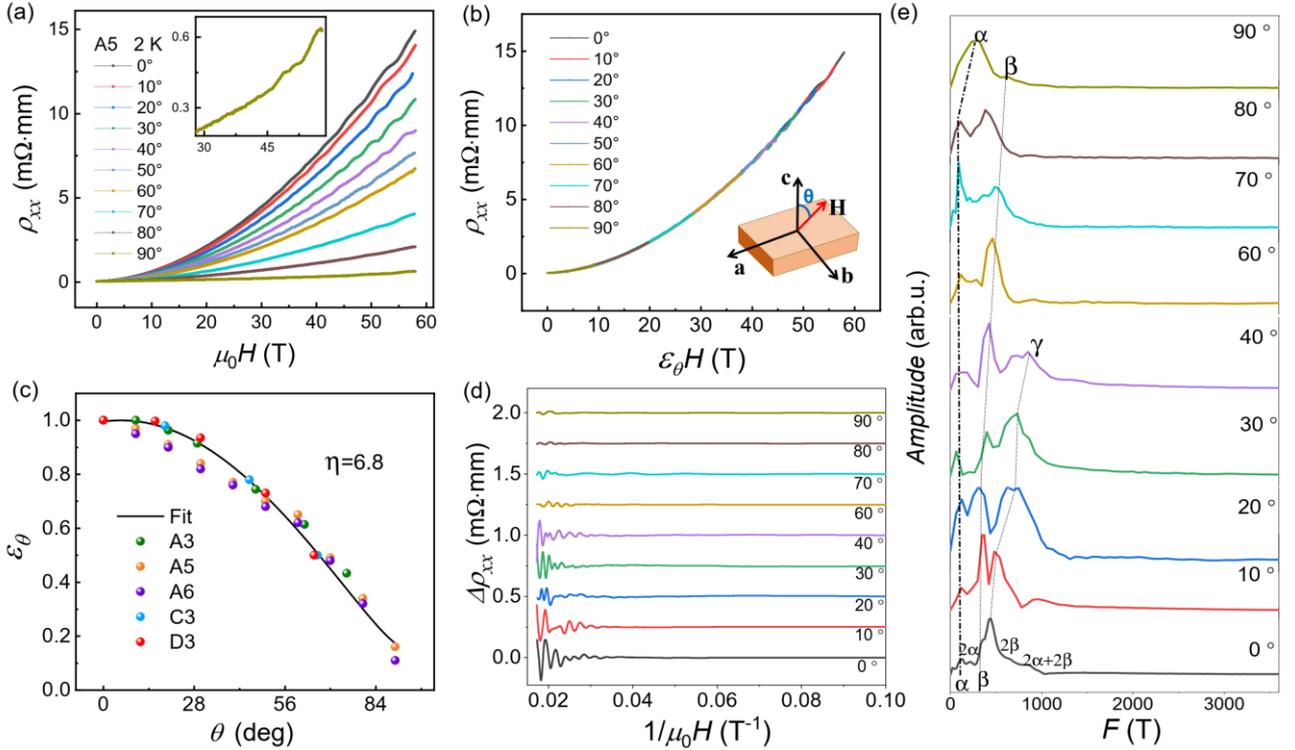

FIG. 4. (a) MR behavior at selected angles with the magnetic field changed from perpendicular ($\theta=0°$) to in-plane ($\theta=90°$). (b) The MR vs magnetic field with $H$ scaled by a factor $\varepsilon_\theta=(\cos^2\theta+\eta^2\sin^2\theta)^{1/2}$ with $\eta$ describing the anisotropy. (c) Angle dependence of $\varepsilon_\theta$ at 2 K. The points are derived from experimental results and the line is the fit with $\varepsilon_\theta=(\cos^2\theta+\eta^2\sin^2\theta)^{1/2}$. (d) The angle-dependent MR oscillations after subtracting a polynomial background. (e) Angular dependent FFT results show the anisotropy of the Fermi pockets. The plots in (d) and (e) are shifted vertically for clarity.

For an in-plane field perpendicular to the current $I$, the HfTe$_2$ shows a MR effect of about 450% at 2 K [Fig. 5(a)]. In the Lorentz-free configuration with $H//I$, the MR is 100% without saturation. The in-plane unsaturated MR may be related to the open-orbit Fermi surface and deserves further study [46,47]. When the direction of the magnetic field is rotated in the $ab$ plane [inset of Fig. 5(a)], AMR and transverse resistivity are observed [Figs. 5(b) and 5(c)]. The phenomenon in topological materials is usually called PHE [48-53]. We calculated the average of the measured Hall data in both negative and positive fields to eliminate the normal Hall contribution arising from a possible out-of-plane filed component in the measurements. The PHE data is further processed by the formula $\rho^{PHE}_{yx} = [\rho_{yx}(\varphi) - \rho_{yx}(\pi-\varphi)]/2$ to exclude the in-plane AMR contribution [18]. The obtained PHE data [Fig. 5(c)] show peaks near 135° (315°) and valleys near 45° (225°) demonstrating a period of 180°. The results are consistent with the expected resistivity tensor considering chiral anomaly [48]: $\rho^{PHE}_{yx} = -\Delta\rho^{chiral}\sin\varphi\cos\varphi$, $\rho_{xx} = \rho_\perp - \Delta\rho^{chiral}\cos^2\varphi$. The $\Delta\rho^{chiral} = \rho_\perp - \rho_{//}$, in which $\rho_\perp$ and $\rho_{//}$ represent the resistivity with the field perpendicular ($\varphi=90°$) and parallel ($\varphi=0°$) to the current, respectively.

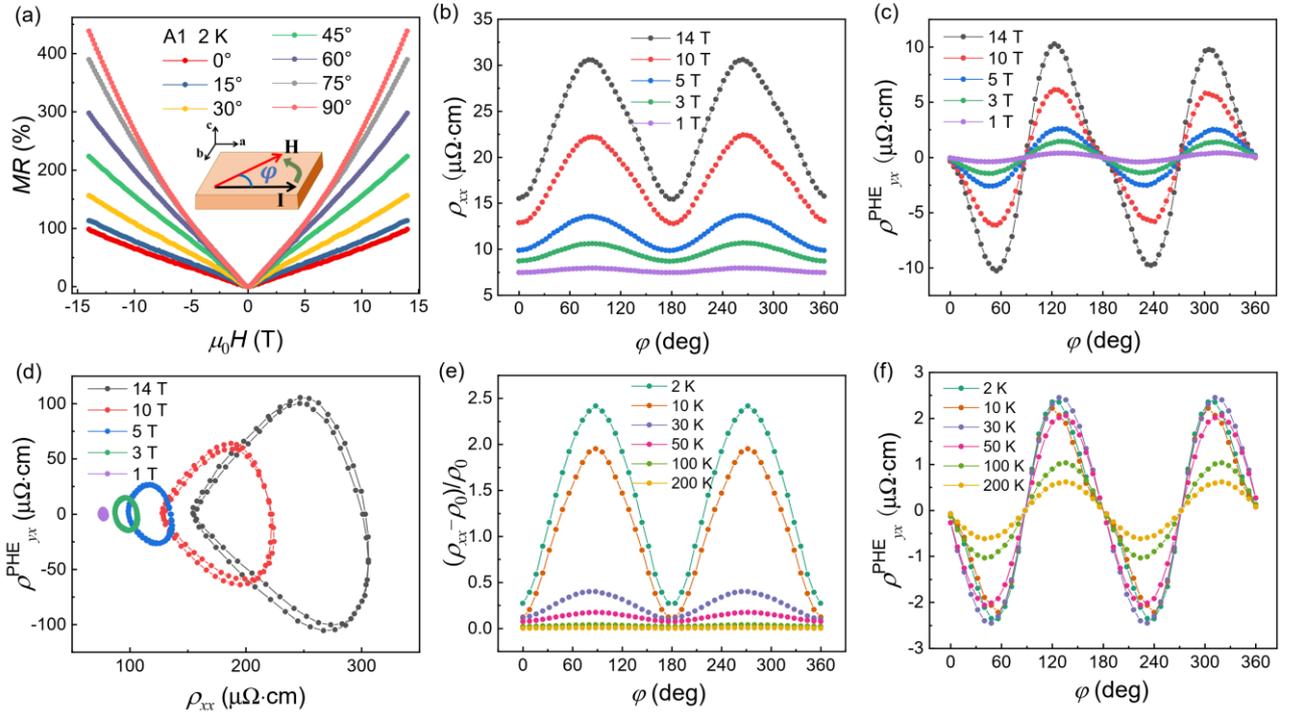

FIG. 5. (a) The in-plane MR curves at different angles ($\varphi=0°$ represents the field parallel to the current). The (b) AMR and (c) PHE at 2 K under different magnetic fields. (d) The orbits obtained by plotting $\rho^{PHE}_{yx}$ and $\rho_{xx}$ with angle $\varphi$ as the parameter. The temperature dependence of (e) AMR and (f) PHE in HfTe$_2$.

A widely used criterion to test the origin of the PHE is to plot the amplitude of PHE vs in-plane AMR with $\varphi$ as a parameter under fixed magnetic fields [Fig. 5(d)]. For the PHE dominated by chiral anomaly, it is believed that the parametric plot pattern of the system should be concentric around the center [54]. Otherwise, the PHE cannot be attributed to the chiral anomaly. The plots of our results are shown in Fig. 5(d), which seems to exclude the chiral anomaly as the origin. In fact, the pattern can be corrected to be concentric around the center if we consider an possible out-of-plane MR contribution in the results. In a system with extremely large MR effect, a small angle misalignment with an out-of-plane field component could induce obvious shift of $\rho_{xx}$ to a larger average value at high magnetic fields. The influence of the perpendicular field MR can change an intrinsic shock-wave pattern to a non-concentric pattern. Thus, we want to point out that the chiral anomaly cannot be excluded for the observation of PHE only by the pattern plot. The absence of negative MR can be an important signature of the absence of chiral anomaly though the negative MR might also be influenced and overwhelmed by a positive background. Since the Fermi level is predicted to be far away from the Dirac point [23,25], it is understandable that the effect of Dirac fermions is not obvious in the transport results. However, it is worth noting that the LK fit on the SdH oscillations suggests a nonzero Berry phase for the larger hole pocket. The nonzero Berry phase can act as the origin of the PHE [53]. On the other hand, the PHE can be closely related to the AMR effect [18]. In this scenario, the temperature dependence of AMR and PHE should be similar. Our results [Figs. 5(e) and 5(f)] indicate that both AMR and PHE are robust to high temperatures. The amplitudes of both AMR and PHE decrease with increased temperature above 30 K, while the amplitude of PHE is almost unchanged at lower temperatures though the AMR amplitude changes. The difference may be a signature that both the nonzero Berry curvature and in-plane AMR contribute to the observed PHE phenomenon.

## IV. CONCLUSIONS

To conclude, we map the Fermi surface of the semimetal $HfTe_2$ by the SdH quantum oscillations and report the PHE phenomenon using the self-Te-flux grown single crystals. The high quality crystal shows a largest MR effect of $1.1 \times 10^4$ % at 14 T and 2 K without saturation. The SdH oscillations identify three Fermi pockets, among which the larger hole Fermi pocket shows signatures of open-orbit characteristic and nontrivial Berry phase. The effective mass, carrier density and mobility are quantified for each pocket. The material exhibits PHE for a wide temperature range and may be attributed to the combined contribution from both the nonzero Berry curvature and the in-plane AMR effect. The results are important for deep understanding of the Fermi surface and the topological nature of $HfTe_2$.

## ACKNOWLEDGMENTS

We acknowledge H. Z. and Y. Z. for the help in the work. This work is financially supported by the National Natural Science Foundation of China (Grant No. 21BAA01133, 12004441, 92165204, 11974431, 11774434, U2130101), the Guangdong Basic and Applied Basic Research Foundation (No. 2023A1515010487), the Guangzhou Basic and Applied Basic Research Foundation (No. 202201011109) and the Fundamental Research Funds for the Central Universities (No. 22hytd07).
Q.L., G.J., and N.T. contributed equally to this work.

Corresponding authors
*wanghch26@mail.sysu.edu.cn
*zuohuakun@163.com